\documentclass[useAMS,usenatbib]{mn2e}

\usepackage{graphics}
\usepackage{epsfig}
\usepackage{longtable}

\title[NGC 2768 Radial Kinematics]
{Radially Extended  Kinematics in the S0 Galaxy NGC 2768 from  
Planetary Nebulae, Globular Clusters and Starlight}
\author[D. A. Forbes et al.]{Duncan A. Forbes$^{1}$\thanks{E-mail:
dforbes@swin.edu.au}, Arianna Cortesi$^{1,2}$, Vincenzo Pota$^{1}$,  Caroline
Foster$^{3}$,  
\newauthor 
Aaron J. Romanowsky$^{4}$, Michael R. Merrifield$^{2}$, Jean
P. Brodie$^{4}$, Jay Strader$^{5}$, 
\newauthor
Lodovico Coccato$^{6}$, Nicola Napolitano$^{7}$
\\
$^{1}$Centre for Astrophysics \& Supercomputing, Swinburne
University, 
Hawthorn VIC 3122, Australia\\
$^{2}$School of Physics and Astronomy, University of Nottingham,
University Park, NG7 2RD Nottingham, UK\\ 
$^{3}$European Southern Observatory, Alonso de Cordova 3107,
Vitacura, 
Santiago, Chile\\
$^{4}$University of California Observatories, Santa Cruz, CA 95064, USA\\
$^{5}$Harvard-Smithsonian Center for Astrophysics, Cambridge, MA 02138, USA\\
$^{6}$European Southern Observatory, Karl-Schwarzschild-Strasse
2, 85748 Garching, Germany\\
$^{7}$Istituto Nazionale di Astrofisica, Observatorio Astronomico
di Capodimonte, Via Moiarello 16, 80131, Naples, Italy 
}
\begin{document}


\pagerange{\pageref{firstpage}--\pageref{lastpage}} \pubyear{2002}

\maketitle

\label{firstpage}

\begin{abstract}
There are only a few tracers available to probe the kinematics of
individual early-type galaxies beyond one effective
radius. Here we directly compare a sample of planetary
nebulae (PNe), globular clusters (GCs) and galaxy starlight velocities 
out to $\sim$4 effective radii, in the S0 galaxy NGC 2768.   
Using a bulge-to-disk decomposition of a K-band image 
we assign PNe and starlight to either the
disk or the bulge. We show that the bulge PNe and bulge starlight
follow the same radial density distribution as the red
subpopulation of GCs, whereas the disk PNe and disk starlight are
distinct components. 
We find good kinematic 
agreement between the three tracers to several effective radii 
(and with stellar data in the inner
regions).  
Further support for the distinct nature of
the two galaxy components come from our kinematic analysis. 
After separating the tracers into bulge and disk components 
we find the bulge to be a slowly rotating pressure-supported
system, whereas the disk reveals a rapidly
rising rotation curve with a declining velocity dispersion
profile. The resulting V$_{rot}$/$\sigma$ ratio for the disk 
resembles that of a spiral galaxy and hints at an
origin for NGC 2768 as a transformed late-type
galaxy. A two-component kinematic analysis for a sample of S0s will help
to elucidate the nature of this class of galaxy. 
\end{abstract}


\section{Introduction}

Based on studies of the morphology-density relation 
in clusters (Dressler 1980) and groups (Wilman et al. 2009),
it has been suggested that lenticular (S0) galaxies  
may be the descendants of spirals that have undergone some
evolutionary process (e.g. ram pressure
stripping, galaxy harrassment, gas starvation and/or
mergers). Recent investigations of S0s have 
studied their metallicity gradients (Bedregal et al. 2011), 
Tully-Fisher relation (Bedregal, 
Aragon-Salamanca \& Merrifield 2006) and stellar populations
(Aragon-Salamanca, Bedregal \& Merrifield  2006). However, 
the origin of S0s is still a subject of much debate (e.g. Kormendy
\& Bender 2012). 
The internal kinematics of galaxies are a key tool to understanding
their structure and formation histories, and S0s are no exception. 
For example, the kinematics will be largely unaffected if S0s
were formerly spirals that have been simply stripped of
gas or if they were involved in a relatively minor merger.

Although half of the
stellar mass within a galaxy lies within one effective radius (R$_e$),
more than 90\% of the {\it total} mass and 
angular momentum does not (Romanowsky \& Fall 2012). Thus in order to
examine the internal kinematics and total mass 
of early-type galaxies one must probe well beyond 1 R$_e$. 
elliptical and S0 (early-type) galaxies often lack the 
significant quantities of extended HI gas commonly found in
spirals, so the kinematic tracers are
the galaxy starlight itself, planetary nebulae
(PNe) and globular clusters (GCs).

Using the underlying starlight of a galaxy to probe its stellar
kinematics is perhaps the preferred method, however 
the surface brightness of a galaxy declines rapidly with
increasing radius so it is very difficult to obtain high quality 
spectra beyond a few effective radii without a large investment of
8m-class telescope time (Coccato et al. 2010) or using deep single
pointings (Weijmans et al. 2009). PNe and
GCs have the advantage that they are ubiquitous in the halos of
early-type galaxies out to large galactocentric
radii (5--10 R$_e$). Although there have been several studies of PNe and GC
system kinematics in early-type galaxies, very few studies have directly
compared them to each other, or to results from galaxy starlight
over a common radial range. 

Luminous PNe are the end product of low mass stars. However, there is still
debate as to whether they arise from normal single-star 
evolutionary processes or from mass transfer in a binary star
system (Ciardullo et al. 2005)
In the former case, the PNe
observed in early-type galaxies would have an age of 
$\sim$1.5 Gyr, and in the latter case they could be as old as 10
Gyr. 
Coccato et al. (2009) showed that the radial surface
density of PNe follows the galaxy starlight in early-type
galaxies and that they are useful probes of galaxy kinematics. 
However, ellipticals with embedded thick disks and S0 galaxies may
contain two subpopulations of PNe, 
one associated with the disk and one with the bulge, as seen for
spiral galaxies (Nolthenius \& Ford 1987; Hurley-Keller et
al. 2004).
Not accounting for the different kinematics of these distinct
PNe subpopulations could lead to misleading results as illustrated
recently by Cortesi et al. (2011) for the lenticular galaxy NGC
1023. Furthermore, Dekel et al. (2005) suggested that an
intermediate-aged population of PNe may 
have `contaminated' the PNe velocity dispersions of some galaxies
in the early-type sample of Romanowsky et al. (2003) and hence
impacted the resulting
mass analysis. 

The globular cluster (GC) systems of all large galaxies, irrespective
of Hubble type, generally consist of two subpopulations -- 
blue (or metal-poor) and red (or metal-rich). 
Both of these
subpopulations are thought to have ages $\ge$ 10 Gyr and hence trace
old stellar populations (for a review of GC system properties see
Brodie \& Strader 2006). 
The blue
subpopulation is associated with galaxy halos (Forte
et al. 2005; Forbes et al. 2012) whereas the red
subpopulation has been shown to share many properties with the
spheroid/bulge stars of early-type galaxies (Strader et al. 2011;
Forbes et al. 2012),  
including their kinematics (Pota et al. 2012). We note that the
association of red GCs with the bulge and not with a thin disk component 
extends to spiral galaxies, including our own  (Minniti
1995; Cote 1999) 
and the Sombrero (Forbes, Brodie \& Larsen 2001).  


To better understand the issues discussed above it is important 
to directly compare different kinematic tracers for the same galaxy.
Here we combine starlight, PNe and GC data for an archetype
lenticular galaxy NGC 2768 and directly 
compare these different kinematic tracers in the same galaxy. 
The galaxy is a  
nearby, near edge-on S0 (Sandage, Tammann \& van den
Bergh 1981), although we note that it was originally classified as an E6
in the RC3 catalogue. 
According to Wikland et al. (1995) is it an isolated galaxy
however it has also been classified as part of the Lyon Group of Galaxies
(LGG) 167 (Garcia 1993).
It reveals ionised gas and a minor axis dust lane (Kim 1989). 
The central ionised gas and stars are known to have different kinematics
(Fried \& Illingworth 1994) suggesting an external origin for the
gas. NGC 2768 is a rare example of an early-type galaxy with
detectable CO emission (Wikland et al. 1995) and the host of a
Calcium-rich supernova type Ib (Filippenko \& Chornock 2000). 
The effective radius of the galaxy is 1.06 arcmin 
(Proctor et al. 2009; Cappellari et al. 2011). For a distance of
21.8 Mpc (Cappellari et al. 2011), this corresponds to 6.7 kpc. 

\section{Kinematic Tracers}

\subsection{Galaxy Starlight Data}

Using a new technique to extract integrated kinematic information
of the
underlying galaxy starlight from a multi-slit spectrograph,
Proctor et al. (2009) presented 2D stellar kinematics for NGC
2768 out to $\sim$3 R$_e$. 
Here we have carried out a re-analysis of the Proctor et
al. galaxy data after re-defining the sky scaling index continuum
passbands to avoid any strong spectral features associated with
the galaxy, as well as any
skylines in order to refine the sky subtraction
(see Foster et al. 2009 for more details). The resulting velocity
and velocity dispersion profiles 
are similar to those published in Proctor et al. (2009).
The total number of positions with stellar kinematics
available are 104 and our full dataset is given in Table 1 of the Appendix.

\subsection{Planetary Nebulae (PNe) Data}

Velocity data for 315 PNe in NGC 2768 were acquired using the PN.S spectrograph
(Douglas et al. 2002) in 2007 and are available at: 
www.strw.leidenuniv.nl/pns/~
Details of the data reduction
procedure and analysis can be found in Cortesi et al. (2011, 2012
in prep.). 
Following Coccato et
al. (2009), a uniform magnitude cut
has been applied and radial incompleteness tests
carried out. Thus each bin of the PNe surface density distribution
is complete to a given magnitude and 
has been statistically corrected for any radial
incompleteness. All are spectroscopically confirmed. 
The PNe data reach out to $\sim$5R$_e$.

\subsection{Globular Cluster (GC) Data}

The radial surface density distribution and velocity data for GCs
comes from the imaging and spectroscopy of Pota et al. (2012).
Briefly, imaging from HST
allows us to model and subtract the galaxy light, and hence 
detect GCs in the galaxy inner regions with little or no radial
incompleteness. The resulting
surface density distribution for over 500 GCs 
is a combination of HST data in the
inner regions and Subaru data in the outer regions, with a
background level subtracted. Blue and red GCs 
were separated according to the local minimum of their bimodal 
colour distribution, i.e. at R--z = 0.56 (Pota et al. 2012).

Follow-up Keck 
spectroscopy returned 112 kinematically-confirmed blue and red GCs (Pota et
al. 2012). The blue GCs, associated with galaxy halos
(Forbes et al. 2012), were found not to rotate (Pota et
al. 2012). 
Here we only consider further the 62 GCs from the red subpopulation.  
For these GCs the mean
velocity is 1353 $\pm$ 3 kms $^{-1}$ 
in excellent agreement with the galaxy systemic
velocity of 1353 km s $^{-1}$ (Cappellari et al. 2011).  
The red GCs reach out to $\sim$4 R$_e$ and we henceforth assume they are 
all associated with the bulge of the galaxy.

\begin{table*}
\caption{Bulge-to-disk decomposition of NGC 2768.
}
\begin{tabular}{lrr}
\hline
Parameter  & Disk & Bulge\\
\hline
R$_d$, R$_e$ (${'}$)& 0.72 & 0.84\\
Sersic n & $\equiv$1 & 4.65\\
K (mag) & 8.19 & 7.23\\
b/a & 0.29 & 0.66\\
\hline
\end{tabular}
\end{table*}

\section{Results and Discussion}

\subsection{Spatial Distributions}

The locations of the 481 starlight, PNe and red GC positions with
kinematic data in NGC 2768
are shown in Figure 1. The kinematic 
data points are well distributed across the surface of the
galaxy. In some cases the GC and starlight velocities come from
the same position on the sky. 

Before examining the kinematic data, we 
compare the 1D radial distribution of the PNe and red GCs
with the galaxy starlight. We use the results of the 
starlight decomposition into bulge and 
disk components from Cortesi et al. (2012, in prep.). The general 
method is to create a model that can be used to assign a probability
that a given position is associated with 
either the disk or bulge component as described in  
Cortesi et al. (2011). 
Briefly, it involves a 2D decomposition
of a K-band image of NGC 2768 from the 2MASS survey. From this, 
parameters for the disk and bulge, assuming an
exponential and Sersic light profile respectively, are obtained. 
The resulting major axis scale lengths, Sersic indices, K-band
magnitudes and axial ratios are given
in Table 1. The 
2D disk/bulge probability map is shown in Figure 1.

The bulge has a position angle that is consistent with the major
axis of the disk. We also find the bulge and disk to have similar 
scale sizes. 
Using a distance of 21.8 Mpc and the K-band magnitudes quoted in
Table 1 (assuming a solar value of K = 3.28)
we can calculate the luminosity in each component. To calculate
masses, we need to assume a mass-to-light ratio that depends on
both age and metallicity. The advantage of working in
the K-band is that it is relatively insensitive to metallicity
and we simply assume solar. 
In a study of age
gradients in S0 galaxies Bedregal et al. (2011) found age 
gradients to be positive (with disks younger than bulges) or
flat. The typical mean age was around 5 Gyr. Although NGC 2768
shows some indication of recent star formation (e.g. a type Ib
supernova in 2000), we assume that a mean age of 5 Gyr is more
appropriate.   
The mass-to-light ratio in the K-band for a near solar 5 Gyr
population is 0.6 (e.g. Forbes et al. 2008). 
This gives the mass of the disk to be 
3.1 $\times$ 10$^{10}$ M$_{\odot}$ and 7.5
$\times$ 10$^{10}$ M$_{\odot}$ for the bulge. Thus the 
bulge-to-total mass ratio is 0.7.

Figure 2 shows the 1D radial surface density distribution of
disk and bulge PNe and the red (bulge) GCs with the galaxy
starlight profiles. The
radial bins, using geometric circular radii, 
are chosen to have similar numbers of objects in each
bin with Poisson errors shown. 
Our assumption of associating the red GCs with the bulge 
is supported by the similarity between their 
surface density profile and that of the bulge starlight. 

\subsection{Kinematic Profiles}

We next explore the radial kinematic (rotation velocity and dispersion) 
profiles of our three
tracers.  For the PNe we follow the
method of Cortesi et al. (2011) which calculates the disk and
bulge kinematics separately in probability-weighted radial
bins, with disk/bulge probabilities taken from our spatial
analysis. A likelihood clipping is
employed to reject data points that are more than $\sim$2$\sigma$ from
either the disk or the bulge model, leaving 289 PNe data points.
The model fits for rotation velocity and velocity
dispersion with only the minimal assumption that the
line-of-sight velocity distributions for the disk
and bulge are drawn from Gaussians. 

Conducting a similar disk/bulge analysis on the red GCs revealed
that none were assigned to the disk with more than 68\%
confidence. This, along with our finding that the
surface density of the red GCs follows that of the bulge
starlight (Figure 2), supports our assumption that the
red GCs probe the bulge component only. Thus for the kinematic
modelling all of the red GCs are assigned to the bulge.

Unlike our two discrete tracers, the starlight data are 
integrated 
quantities for which the rotation velocity
and velocity dispersion come directly from the
measurements. 
Ideally, with full wavelength coverage and a high
signal-to-noise 2D spectrum we could attempt to decompose the
starlight into disk and bulge components (Johnston et al. 2012).
However for our data, which are modest signal-to-noise data 
from a restricted wavelength range
around the Ca Triplet lines, we adopt a simpler approach.
We assign each
starlight position to {\it either} the disk or the bulge based on
the probability map of Figure 1. The rotation amplitude is fit by
an inclined disk model rotating along the major axis using equation 
3 from Foster et al. (2011) which fits the rotation velocity and
velocity dispersion simultaneously. 

In Figure 3 we show how the
starlight and PNe rotation velocity depends on the disk/bulge
probability. In both cases the rotation velocity declines smoothly from the
disk-dominated regime  to the
bulge-dominated regime. We note that the
velocity dispersion (not shown) does not depend strongly on the disk/bulge
probability. The figure also
gives the azimuthal dependence on the probability. This
shows that low (disk-like) probabilities are exclusively associated with
position angles about the 
major axis (90$^{o}$ and
270$^{o}$) as expected. For the starlight data 
we have chosen a probability cut of
63\%. This combined with a restriction of velocity errors 
to be less than 50 km s$^{-1}$ gives a final sample of 
23 disk and 22 bulge-dominated starlight data points. We
use these subsamples to calculate the starlight disk and bulge
kinematics but recognise that they are still `contaminated'
somewhat by the other component. To
explore the sensitivity of this cut, and hence the contamination
effect, we have varied the probability
by 9\% either side of our chosen cut. 
The resulting systematic change in disk and
bulge starlight kinematics are included in our errors.


In Figures 4 and 5 we show the disk and bulge kinematic 
profiles from our two-component decomposition. Each radial bin is
chosen to have similar numbers in each bin with 289 (disk and
bulge) PNe in 6 radial bins, 122 red (bulge only) GC in 2 bins and 45 (disk
and bulge) starlight data points in 2 bins. As detailed above the
rotation velocity and velocity dispersion in each radial bin are
computed following Cortesi et al. (2011) for the PNe and red GCs,
and Foster et al. (2011) for the starlight data. 
For the stellar velocity dispersion we show a continuous profile. 
We also include the
inner region stellar kinematic profiles from the SAURON
instrument (Emsellem et al. 2004). For the latter we have simply extracted
profiles $\pm 30{^o}$ about the major and 
$\pm 60{^o}$ about the minor axes to
approximate the disk and bulge profiles respectively. 
The resulting disk
and bulge data are both plotted 
as a function of a generalised radius, i.e.\\

\noindent
R = $\sqrt{R_{maj}^2 q + R_{min}^2/q}$, \hspace{1.6in} (1)\\

\noindent
where $R_{maj}$ is the semi-major axis, $R_{min}$ is the
semi-minor axis, and q is the axial ratio of 0.46 from 2MASS.

Figures 4 and 5 shows generally good agreement 
between the PNe kinematics
and the kinematics probed by red GCs and direct starlight to
large radii for NGC 2768.
We also see reasonable consistency in the inner regions with the
SAURON stellar kinematic data. 
We support the earlier work by Cortesi et al. (2011), on the S0
galaxy NGC 1023, finding evidence for distinct populations of 
disk and bulge PNe.

The disk component reveals a rapid rise reaching a maximum
velocity of $\sim$ 270 km s$^{-1}$ at a few scale lengths. 
We note that the true circular velocity may be closer to 311 km
s$^{-1}$ given by Davis et al. (2011) from their modelling of molecular
gas kinematics which is unaffected by the asymmetric drift that may
influence the kinematics derived from stars. 
Beyond the very inner regions, we find a continuously declining 
velocity dispersion profile to very low values. 
In a sample of late-type spiral galaxies, Bottema (1993) found 
the velocity dispersion to decline exponentially
with radius reaching very low values at a few scale lengths. 
The bulge reveals mild rotation with a  
near constant velocity of $\sim$ 100 km s$^{-1}$  
and a mean velocity dispersion of $\sim$120 km s$^{-1}$ at large
radii.

Figure 4 also shows that V$_{rot}$/$\sigma$ for the disk rises
continuously with radius.
Such a trend suggests that NGC 2768 could be
descended from a galaxy with an extended disk. 
The  simulations of Bournaud et al. (2005) showed that  
such high values of V$_{rot}$/$\sigma$ are maintained after  
a minor merger. Minor mergers may also 
help to build-up the bulge mass.  
We note that an isotropic oblate rotator of b/a = 0.66 (see Table 1) would be
expected to have V$_{rot}$/$\sigma$ $\sim$ 0.7 (Davies et
al. 1983), similar to our average value. 
The bulge of NGC 2768 (Figure 5) resembles a pressure-supported 
system similar to those seen
for the bulges of other 
spiral galaxies (MacArthur, Gonzalez \& Courteau 2009).
Thus we find NGC 2768 to have kinematic properties that resemble
those of a spiral galaxy (although with a
dominant bulge).  We note that 
determining the exact 
differences and testing the various evolutionary processes
suggested for S0
formation would 
require a detailed analysis beyond the scope of this paper.

\section{Conclusions}

By combining galaxy starlight, PNe, and red GC data for NGC 2768 we
have shown that the red subpopulation of GCs and a subpopulation
of PNe follow the same radial surface density profile as the
bulge component of the galaxy starlight. An additional distinct,
and presumably younger, 
PNe subpopulation is associated with the galaxy disk. 

We have also investigated the radially extended kinematics of these
three tracers out to $\sim$ 4R$_e$ and show that they are in very
good agreement with each other over a large range of radii. 
Using a 2D probability map generated from a K-band bulge-to-disk
decomposition we estimate the bulge-to-total ratio to be 0.7, with both
components having a similar scale size.  
Using this two-component 
decomposition we examined the kinematics of the bulge and
disk separately finding that the disk component
reveals rapid rotation and a falling velocity
dispersion profile resulting in an increasing V$_{rot}$/$\sigma$ ratio
with radius. Such
a behaviour is similar to that seen in the disks of late-type spiral galaxies.
We also found a slowly rotating bulge, indicative of a
pressure-supported system.

Although NGC 2768 is an S0 galaxy, many elliptical galaxies
contain embedded disks and potentially an associated subpopulation of
disk PNe. This work shows that bulge and disk components may need to
be taken into account in any kinematic, or mass, analysis 
of early-type galaxies. This is particularly true when using
kinematics to help unravel the origin of S0 galaxies.

\section*{Acknowledgments}

We thank C. Blom, J. Arnold, L. Spitler for their
help with the Keck observations. We thank the referee for several
useful comments. 
CF acknowledges co-funding under the Marie Curie Actions of the 
European Commission (FP7-COFUND).
LC has received funding from the European Community's 
Seventh Framework Programme (/FP7/2007-2013/) under grant agreement No 229517.
JB. and AR  acknowledge support from the NSF through grants
AST-0808099, AST-0909237, and AST-1109878.

\begin{figure*}
\begin{center}
\includegraphics[angle=0,scale=0.5]{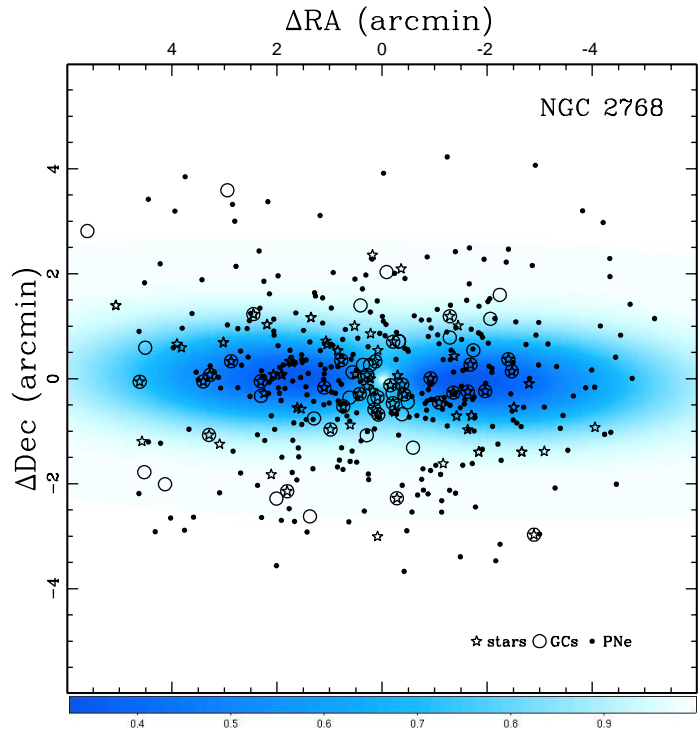}
\caption{Distribution of planetary nebulae (PNe, solid dots), 
red globular clusters (GCs, open circles) and starlight
positions (stars) for which
velocity information is available for NGC 2768. Also shown is 
a greyscale
corresponding to the probability of association with the bulge
component 
(high values in the greyscale bar shown below). 
The effective radius of the galaxy is 1.06 arcmin.
}
\end{center}
\end{figure*}

\begin{figure*}
\begin{center}
\includegraphics[angle=0,scale=0.5]{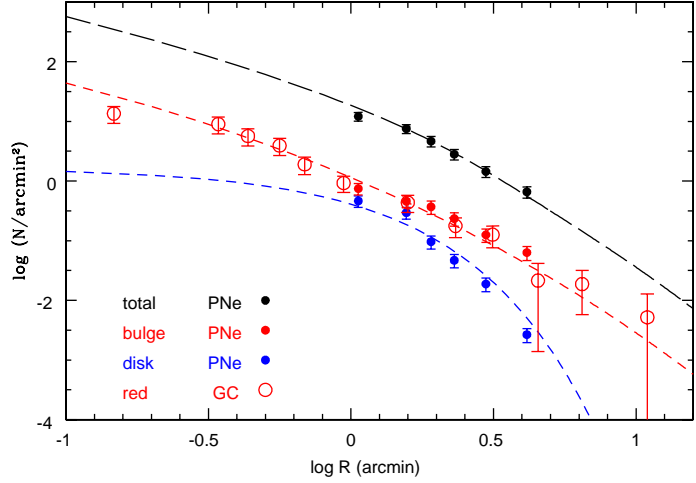}
\caption{Surface density of planetary nebulae (PNe) and 
globular clusters (GCs) in NGC 2768 as a function of galaxy major axis. 
The surface density of red GCs 
are shown as red open circles. The surface density
of PNe assigned to the disk and bulge components are shown by
blue and red dots respectively. 
The combined total PNe surface density is given by black dots. 
The disk, bulge and total galaxy starlight profiles from a
bulge-to-disk decomposition are given by
blue, red and black dashed lines. The normalisation in the
vertical direction is arbitrary (i.e. the disk and bulge
distributions have been offset for clarity; in reality the bulge
dominates at small and large radii). 
A good correspondence is
seen between the bulge PNe, bulge starlight and red GCs. 
[This plot is best viewed in colour.]
}
\end{center}
\end{figure*}

\begin{figure*}
\begin{center}
\includegraphics[angle=0,scale=0.6]{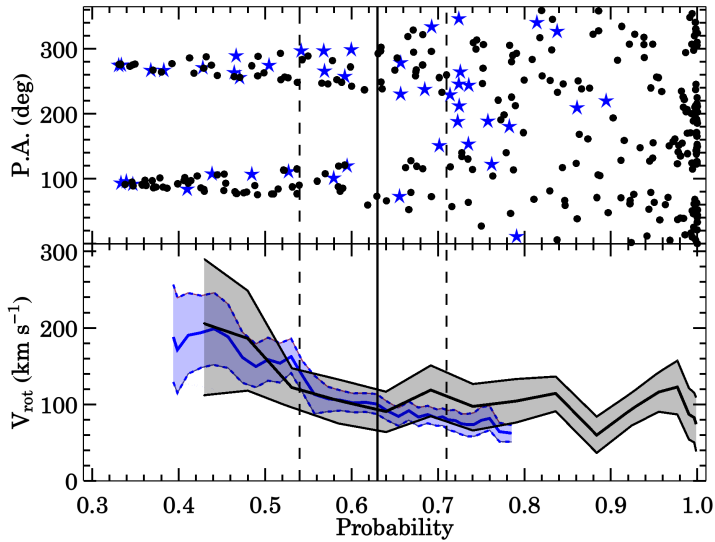}
\caption{Stellar and PNe 
rotation velocity of NGC 2768 as a function of disk/bulge
  probability. The probability associated with the disk (low
  values) or the bulge (high values) come from our bulge-to-disk
  decomposition of a K-band image (see Section 3.1). 
The {\it upper} panel shows the position angles of
  the starlight (stars) and PNe (solid dots) data points, 
with the galaxy major axis located
  at $\sim90^{o}$ and 270$^{o}$. The {\it lower} panel shows
  a rolling average of the stellar (blue line) and PNe (black
  line) rotation velocities and the
  1$\sigma$ confidence limits. The vertical lines at a
  probability of 0.63 and $\pm$0.09 indicate our chosen separation between
  disk ($<$0.63) and bulge
  ($>$0.63) for the starlight data, 
and the range of values that we explore for
  systematic effects. [This plot is
  best viewed in colour.]  
}
\end{center}
\end{figure*}

\begin{figure*}
\begin{center}
\includegraphics[angle=0,scale=0.6]{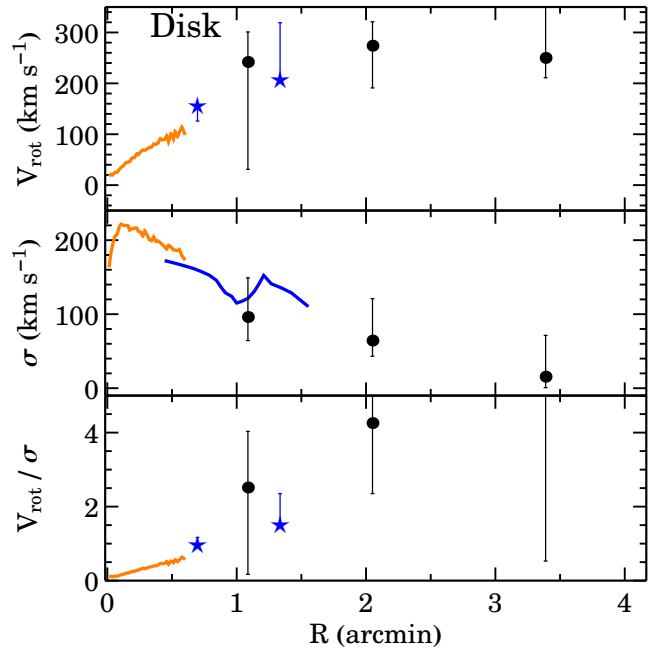}
\caption{Disk kinematic profiles for NGC 2768. Symbols are solid
  dots for PNe, open circles for GCs and blue stars for starlight
  data. The stellar velocity dispersion is shown
  as a continuous blue line. 
The continuous orange lines at small radii show the SAURON data along the major
(disk) and minor (bulge) axes. 
The {\it upper} panels show the rotation velocity (V$_{rot}$),
the {\it middle} panels
the velocity dispersion ($\sigma$) and the {\it lower} panels the V$_{rot}$/$\sigma$
ratio (note the different scales on the {\it left} and {\it
  right} sides). The outermost disk V$_{rot}$/$\sigma$ data point
is off scale. 
The x axis shows the generalised radius as given by Eq. 1. 
Generally good
agreement is seen between the different kinematic tracers. 
[This plot is best viewed in colour.] 
}
\end{center}
\end{figure*}

\begin{figure*}
\begin{center}
\includegraphics[angle=0,scale=0.6]{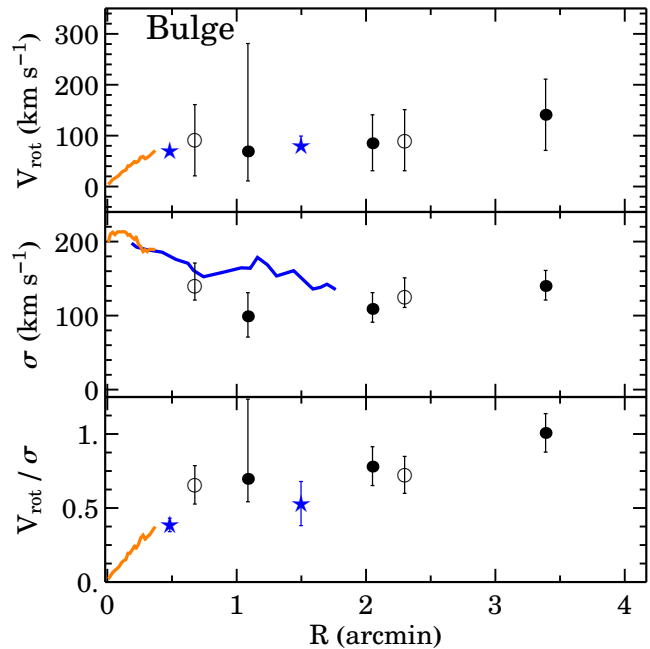}
\caption{Bulge kinematic profiles for NGC 2768. Symbols are the
  same as Figure 4.
The continuous orange lines at small radii show the SAURON data along the major
(disk) and minor (bulge) axes. 
The {\it upper} panels show the rotation velocity (V$_{rot}$),
the {\it middle} panels
the velocity dispersion ($\sigma$) and the {\it lower} panels the V$_{rot}$/$\sigma$
ratio (note the different scales on the {\it left} and {\it
  right} sides). 
The x axis shows the generalised radius as given by Eq. 1. 
Generally good
agreement is seen between the different kinematic tracers. 
[This plot is best viewed in colour.] 
}
\end{center}
\end{figure*}

\newpage 


\setcounter{table}{0}

\begin{table*}
\begin{center}
\caption{Galaxy kinematic data. Columns 1 and 2 give the position in Right Ascension and Declination (J2000), respectively. Columns 3 and 4 give the galactocentric radius. The measured kinematic moments are given in Columns 5 to 8. Data for two individual masks are included (Y and Z), however mask Y has been observed twice independently and is labelled Y1 and Y2.}
\begin{tabular}{cccccccc}
\hline
$\alpha$ & $\delta$ & $r$ & PA & $V_{\rm obs}$ & $\sigma$ & $h_3$ & $h_4$\\
(hh:mm:ss)&($^o$:':")&(arcsec)&(degree)&(km s$^{-1}$)&(km s$^{-1}$)&&\\
(1)&(2)&(3)&(4)&(5)&(6)&(7)&(8)\\
\hline
&&&&Mask Y1&&&\\
\hline
09:11:28.703 & 60:02:48.06 &  92.5 & 297 & 1205$\pm$ 11 & 143$\pm$ 18 &  0.08$\pm$0.09 & -0.02$\pm$0.12\\
09:11:39.886 & 60:01:37.87 &  81.5 & 153 & 1343$\pm$ 14 & 178$\pm$ 19 & -0.02$\pm$0.06 &  0.02$\pm$0.09\\
09:11:20.931 & 60:02:07.85 & 126.4 & 267 & 1148$\pm$  8 & 126$\pm$ 12 &  0.01$\pm$0.07 & -0.12$\pm$0.13\\
09:11:23.223 & 60:02:24.78 & 106.9 & 275 & 1122$\pm$  7 & 107$\pm$ 18 &  0.07$\pm$0.07 &  0.10$\pm$0.08\\
09:11:33.407 & 60:02:16.73 &  29.9 & 273 & 1228$\pm$  5 & 171$\pm$  8 &  0.02$\pm$0.02 &  0.01$\pm$0.02\\
09:11:25.877 & 60:03:16.79 & 152.8 & 306 & 1209$\pm$ 55 & 110$\pm$ 65 &  0.05$\pm$0.09 & -0.03$\pm$0.13\\
09:11:24.979 & 60:01:32.49 & 137.4 & 245 & 1218$\pm$ 15 & 113$\pm$ 20 &  0.06$\pm$0.07 & -0.08$\pm$0.10\\
09:11:22.204 & 60:02:08.74 & 116.5 & 267 & 1127$\pm$  9 &  96$\pm$ 16 &  0.10$\pm$0.08 &  0.05$\pm$0.13\\
09:11:28.835 & 60:01:58.20 &  77.1 & 255 & 1217$\pm$  8 & 142$\pm$ 10 &  0.03$\pm$0.05 &  0.01$\pm$0.05\\
09:11:30.367 & 60:02:41.56 &  74.4 & 297 & 1240$\pm$  7 & 130$\pm$ 11 & -0.02$\pm$0.06 & -0.05$\pm$0.06\\
09:11:19.758 & 60:01:45.71 & 152.5 & 257 & 1197$\pm$ 13 & 123$\pm$ 11 & -0.01$\pm$0.05 & -0.06$\pm$0.07\\
09:11:21.916 & 60:00:13.79 & 297.6 & 224 & 1306$\pm$123 & 211$\pm$117 &  0.01$\pm$0.07 & -0.02$\pm$0.10\\
09:11:38.854 & 60:02:07.47 &  19.2 & 125 & 1376$\pm$  5 & 196$\pm$  8 &  0.01$\pm$0.01 & -0.02$\pm$0.03\\
09:11:36.253 & 60:01:59.28 &  36.3 & 208 & 1309$\pm$  5 & 180$\pm$  8 &  0.00$\pm$0.02 &  0.02$\pm$0.02\\
09:11:37.409 & 60:01:32.92 &  91.8 & 180 & 1314$\pm$ 22 & 212$\pm$ 26 & -0.12$\pm$0.08 &  0.07$\pm$0.17\\
09:11:35.694 & 60:02:12.86 &  14.2 & 260 & 1270$\pm$  5 & 203$\pm$  8 &  0.03$\pm$0.01 &  0.01$\pm$0.02\\
09:11:19.269 & 60:03:04.43 & 165.7 & 290 & 1152$\pm$ 70 &  65$\pm$ 84 &  0.08$\pm$0.10 &  0.01$\pm$0.16\\
09:11:41.096 & 60:01:58.14 &  44.3 & 122 & 1400$\pm$  5 & 174$\pm$  8 & -0.04$\pm$0.02 &  0.02$\pm$0.06\\
09:11:24.573 & 59 59 56.42 & 324.5 & 215 & 1132$\pm$ 85 & 106$\pm$ 54 &  0.00$\pm$0.06 & -0.03$\pm$0.06\\
09:11:34.314 & 60:01:54.53 &  52.3 & 228 & 1274$\pm$  5 & 155$\pm$  8 &  0.00$\pm$0.02 &  0.01$\pm$0.03\\
09:11:39.822 & 60:02:19.25 &  21.0 &  77 & 1403$\pm$  5 & 194$\pm$  8 & -0.02$\pm$0.01 &  0.00$\pm$0.02\\
09:11:31.048 & 60:02:31.14 &  56.5 & 289 & 1221$\pm$  6 & 172$\pm$  9 &  0.04$\pm$0.04 &  0.05$\pm$0.04\\
09:11:36.752 & 60:01:38.21 &  80.9 & 187 & 1329$\pm$ 16 & 184$\pm$ 20 & -0.08$\pm$0.07 &  0.02$\pm$0.12\\
09:11:19.078 & 60:01:59.02 & 145.3 & 263 & 1136$\pm$ 13 &  91$\pm$ 28 &  0.17$\pm$0.08 &  0.00$\pm$0.12\\
09:12:10.561 & 60:01:39.79 & 254.6 &  98 & 1329$\pm$144 & 170$\pm$144 &  0.02$\pm$0.08 & -0.02$\pm$0.12\\
09:12:02.699 & 59 59:31.30 & 388.4 & 131 & 1325$\pm$164 & 228$\pm$153 &  0.01$\pm$0.11 &  0.00$\pm$0.19\\
09:11:59.927 & 60:02:23.17 & 172.0 &  87 & 1560$\pm$ 53 & 131$\pm$102 & -0.01$\pm$0.10 &  0.05$\pm$0.12\\
09:11:59.744 & 60:01:04.39 & 217.6 & 113 & 1481$\pm$ 79 & 164$\pm$107 & -0.05$\pm$0.06 & -0.01$\pm$0.05\\
09:11:57.921 & 60:01:54.50 & 157.0 &  98 & 1500$\pm$ 27 & 149$\pm$ 63 & -0.12$\pm$0.11 &  0.03$\pm$0.12\\
09:11:57.142 & 60:02:35.58 & 158.9 &  82 & 1506$\pm$ 45 & 125$\pm$ 86 & -0.09$\pm$0.10 & -0.05$\pm$0.12\\
09:11:56.449 & 60:02:48.83 & 166.7 &  77 & 1523$\pm$ 47 & 148$\pm$ 87 & -0.01$\pm$0.12 & -0.07$\pm$0.11\\
09:11:53.358 & 60:02:10.14 & 119.7 &  92 & 1523$\pm$ 11 & 127$\pm$ 15 & -0.14$\pm$0.06 & -0.07$\pm$0.12\\
09:11:53.195 & 60:01:00.13 & 193.0 & 122 & 1289$\pm$ 41 & 168$\pm$ 65 & -0.01$\pm$0.06 &  0.00$\pm$0.04\\
09:11:51.588 & 60:01:40.84 & 124.7 & 108 & 1445$\pm$ 19 & 159$\pm$ 18 & -0.18$\pm$0.05 & -0.09$\pm$0.11\\
09:11:50.723 & 60:02:07.76 & 100.2 &  94 & 1500$\pm$  9 & 141$\pm$ 12 & -0.07$\pm$0.04 &  0.01$\pm$0.09\\
09:11:49.342 & 60:01:38.98 & 114.1 & 112 & 1456$\pm$ 15 & 175$\pm$ 16 & -0.11$\pm$0.04 &  0.00$\pm$0.10\\
09:11:48.534 & 60:02:05.28 &  84.7 &  97 & 1490$\pm$  8 & 155$\pm$  8 & -0.11$\pm$0.03 & -0.06$\pm$0.07\\
09:11:48.180 & 60:02:46.95 & 111.1 &  69 & 1460$\pm$ 22 & 141$\pm$ 57 &  0.04$\pm$0.07 &  0.08$\pm$0.08\\
09:11:46.500 & 60:01:52.47 &  80.8 & 108 & 1457$\pm$  7 & 152$\pm$  8 & -0.05$\pm$0.03 & -0.09$\pm$0.05\\
09:11:44.879 & 60:02:19.45 &  57.9 &  86 & 1463$\pm$  6 & 167$\pm$  8 & -0.06$\pm$0.02 & -0.04$\pm$0.04\\
09:11:17.066 & 60:03:14.92 & 191.6 & 291 & 1131$\pm$102 &  33$\pm$114 &  0.03$\pm$0.09 & -0.03$\pm$0.11\\
09:11:14.309 & 60:02:18.90 & 173.3 & 271 & 1182$\pm$ 20 &  98$\pm$ 31 &  0.10$\pm$0.07 & -0.03$\pm$0.08\\
09:11:13.660 & 60:00:43.19 & 279.7 & 243 & 1163$\pm$ 97 &  92$\pm$ 90 &  0.03$\pm$0.06 & -0.02$\pm$0.12\\
09:11:12.873 & 60:02:39.49 & 187.5 & 278 & 1109$\pm$ 40 &  34$\pm$ 60 &  0.02$\pm$0.07 & -0.01$\pm$0.06\\
09:11:11.932 & 60:00:52.67 & 273.9 & 247 & 1202$\pm$ 76 &  76$\pm$ 84 &  0.01$\pm$0.07 & -0.04$\pm$0.10\\
09:11:11.256 & 60:02:01.23 & 201.8 & 266 & 1200$\pm$ 34 &  80$\pm$ 40 &  0.04$\pm$0.06 & -0.02$\pm$0.07\\
09:11:10.340 & 60:01:53.13 & 213.4 & 264 & 1187$\pm$ 55 &  92$\pm$ 56 &  0.03$\pm$0.07 & -0.03$\pm$0.11\\
09:11:01.815 & 60:00:38.63 & 354.6 & 250 & 1175$\pm$116 &  81$\pm$107 &  0.01$\pm$0.07 & -0.03$\pm$0.12\\
09:11:00.672 & 60:01:46.55 & 288.8 & 264 & 1224$\pm$ 93 &  69$\pm$ 84 & -0.01$\pm$0.02 & -0.01$\pm$0.03\\
09:12:03.148 & 60:01:07.31 & 232.9 & 109 & 1406$\pm$ 66 & 247$\pm$ 79 & -0.06$\pm$0.07 & -0.05$\pm$0.07\\
09:10:56.037 & 60:03:10.99 & 323.9 & 280 & 1194$\pm$130 &  68$\pm$117 &  0.01$\pm$0.05 & -0.02$\pm$0.06\\
\hline
&&&&Mask Y2&&&\\
\hline
09:11:28.703 & 60:02:48.06 &  92.5 & 297 & 1209$\pm$ 38 & 185$\pm$ 48 &  0.00$\pm$0.14 &  0.09$\pm$0.14\\
09:11:39.886 & 60:01:37.87 &  81.5 & 153 & 1322$\pm$ 47 & 191$\pm$ 57 & -0.10$\pm$0.12 &  0.10$\pm$0.27\\
09:11:20.931 & 60:02:07.85 & 126.4 & 267 & 1140$\pm$ 25 & 110$\pm$ 45 &  0.05$\pm$0.13 &  0.11$\pm$0.12\\
09:11:23.223 & 60:02:24.78 & 106.9 & 275 & 1127$\pm$ 25 & 132$\pm$ 45 &  0.08$\pm$0.12 &  0.08$\pm$0.11\\
09:11:33.407 & 60:02:16.73 &  29.9 & 273 & 1232$\pm$  5 & 179$\pm$  8 &  0.00$\pm$0.04 &  0.00$\pm$0.03\\
09:11:25.877 & 60:03:16.79 & 152.8 & 306 & 1217$\pm$113 & 151$\pm$117 &  0.05$\pm$0.11 & -0.03$\pm$0.14\\
\hline
\end{tabular}
\label{table:GC}
\end{center}
\end{table*}	

 \begin{table*}
\begin{center}
\contcaption{}
\begin{tabular}{cccccccc}
\hline
$\alpha$ & $\delta$ & $r$ & PA & $V_{\rm obs}$ & $\sigma$ & $h_3$ & $h_4$\\
(hh:mm:ss)&($^o$:':")&(arcsec)&(degree)&(km s$^{-1}$)&(km s$^{-1}$)&&\\
(1)&(2)&(3)&(4)&(5)&(6)&(7)&(8)\\
\hline
09:11:24.979 & 60:01:32.49 & 137.4 & 245 & 1200$\pm$ 98 & 244$\pm$ 743 & -0.15$\pm$0.12 &  0.07$\pm$0.12\\
09:11:22.204 & 60:02:08.74 & 116.5 & 267 & 1127$\pm$ 27 & 102$\pm$ 51 &  0.08$\pm$0.14 &  0.08$\pm$0.13\\
09:11:28.835 & 60:01:58.20 &  77.1 & 255 & 1216$\pm$ 25 & 171$\pm$ 29 & -0.06$\pm$0.10 & -0.03$\pm$0.09\\
09:11:30.367 & 60:02:41.56 &  74.4 & 297 & 1227$\pm$ 19 & 157$\pm$ 23 &  0.01$\pm$0.11 & -0.02$\pm$0.09\\
09:11:19.758 & 60:01:45.71 & 152.5 & 257 & 1187$\pm$ 60 & 145$\pm$ 66 & -0.02$\pm$0.10 & -0.04$\pm$0.10\\
09:11:38.854 & 60:02:07.47 &  19.2 & 125 & 1378$\pm$  5 & 210$\pm$  8 &  0.00$\pm$0.02 &  0.04$\pm$0.05\\
09:11:36.253 & 60:01:59.28 &  36.3 & 208 & 1310$\pm$  6 & 184$\pm$  9 & -0.03$\pm$0.03 &  0.03$\pm$0.07\\
09:11:37.409 & 60:01:32.92 &  91.8 & 180 & 1293$\pm$144 & 321$\pm$113 & -0.10$\pm$0.12 &  0.20$\pm$0.19\\
09:11:35.694 & 60:02:12.86 &  14.2 & 260 & 1273$\pm$  5 & 203$\pm$  8 &  0.01$\pm$0.01 & -0.03$\pm$0.05\\
09:11:19.269 & 60:03:04.43 & 165.7 & 290 & 1148$\pm$143 &  55$\pm$119 &  0.12$\pm$0.11 &  0.00$\pm$0.14\\
09:11:41.096 & 60:01:58.14 &  44.3 & 122 & 1400$\pm$  8 & 169$\pm$ 15 & -0.05$\pm$0.04 &  0.01$\pm$0.08\\
09:11:34.314 & 60:01:54.53 &  52.3 & 228 & 1274$\pm$ 10 & 176$\pm$ 12 & -0.04$\pm$0.06 &  0.04$\pm$0.11\\
09:11:39.822 & 60:02:19.25 &  21.0 &  77 & 1404$\pm$  5 & 190$\pm$  8 & -0.03$\pm$0.02 & -0.01$\pm$0.03\\
09:11:31.048 & 60:02:31.14 &  56.5 & 289 & 1226$\pm$ 13 & 166$\pm$ 15 &  0.03$\pm$0.08 & -0.01$\pm$0.07\\
09:11:36.752 & 60:01:38.21 &  80.9 & 187 & 1311$\pm$ 73 & 300$\pm$ 76 & -0.06$\pm$0.10 &  0.19$\pm$0.19\\
09:11:19.078 & 60:01:59.02 & 145.3 & 263 & 1149$\pm$ 38 &  82$\pm$ 64 &  0.03$\pm$0.13 & -0.01$\pm$0.14\\
09:11:59.927 & 60:02:23.17 & 172.0 &  87 & 1523$\pm$153 &  96$\pm$129 &  0.04$\pm$0.11 & -0.09$\pm$0.11\\
09:11:59.744 & 60:01:04.39 & 217.6 & 113 & 1346$\pm$148 & 148$\pm$117 & -0.01$\pm$0.08 &  0.01$\pm$0.07\\
09:11:57.921 & 60:01:54.50 & 157.0 &  98 & 1488$\pm$ 85 & 172$\pm$118 & -0.14$\pm$0.13 &  0.05$\pm$0.16\\
09:11:57.142 & 60:02:35.58 & 158.9 &  82 & 1481$\pm$ 91 & 149$\pm$144 &  0.10$\pm$0.12 & -0.10$\pm$0.12\\
09:11:56.449 & 60:02:48.83 & 166.7 &  77 & 1621$\pm$160 & 223$\pm$114 &  0.10$\pm$0.08 & -0.01$\pm$0.09\\
09:11:53.358 & 60:02:10.14 & 119.7 &  92 & 1526$\pm$ 34 & 114$\pm$ 69 & -0.12$\pm$0.14 & -0.14$\pm$0.20\\
09:11:53.195 & 60:01:00.13 & 193.0 & 122 & 1398$\pm$125 & 150$\pm$118 &  0.03$\pm$0.04 & -0.01$\pm$0.03\\
09:11:51.588 & 60:01:40.84 & 124.7 & 108 & 1466$\pm$ 34 & 148$\pm$ 58 & -0.12$\pm$0.11 &  0.02$\pm$0.22\\
09:11:50.723 & 60:02:07.76 & 100.2 &  94 & 1477$\pm$ 25 & 286$\pm$ 49 & -0.11$\pm$0.10 &  0.25$\pm$0.38\\
09:11:49.342 & 60:01:38.98 & 114.1 & 112 & 1437$\pm$ 25 & 160$\pm$ 43 & -0.09$\pm$0.09 &  0.04$\pm$0.16\\
09:11:48.534 & 60:02:05.28 &  84.7 &  97 & 1485$\pm$ 20 & 166$\pm$ 36 & -0.12$\pm$0.08 &  0.00$\pm$0.18\\
09:11:48.180 & 60:02:46.95 & 111.1 &  69 & 1457$\pm$ 40 & 197$\pm$ 75 &  0.08$\pm$0.10 &  0.07$\pm$0.09\\
09:11:46.500 & 60:01:52.47 &  80.8 & 108 & 1465$\pm$ 16 & 155$\pm$ 33 & -0.08$\pm$0.07 &  0.03$\pm$0.15\\
09:11:44.879 & 60:02:19.45 &  57.9 &  86 & 1460$\pm$  9 & 172$\pm$ 18 & -0.05$\pm$0.05 &  0.00$\pm$0.08\\
09:11:17.066 & 60:03:14.92 & 191.6 & 291 & 1264$\pm$126 & 118$\pm$144 &  0.02$\pm$0.10 & -0.05$\pm$0.12\\
09:11:14.309 & 60:02:18.90 & 173.3 & 271 & 1237$\pm$ 82 & 225$\pm$ 76 &  0.11$\pm$0.09 &  0.03$\pm$0.12\\
09:11:12.873 & 60:02:39.49 & 187.5 & 278 & 1197$\pm$128 & 122$\pm$120 &  0.04$\pm$0.09 &  0.02$\pm$0.11\\
09:11:11.256 & 60:02:01.23 & 201.8 & 266 & 1132$\pm$115 &  28$\pm$122 & -0.01$\pm$0.04 & -0.01$\pm$0.03\\
09:11:10.340 & 60:01:53.13 & 213.4 & 264 & 1215$\pm$ 39 & 123$\pm$ 60 &  0.01$\pm$0.07 & -0.03$\pm$0.07\\
09:11:05.861 & 60:02:33.21 & 237.2 & 274 & 1099$\pm$170 & 135$\pm$107 & -0.05$\pm$0.08 &  0.01$\pm$0.10\\
09:12:35.618 & 59 58:53.58 & 591.3 & 115 & 1382$\pm$ 99 &  38$\pm$102 & -0.01$\pm$0.03 & -0.02$\pm$0.03\\
09:11:06.573 & 60:02:29.59 & 231.4 & 274 & 1173$\pm$173 &  49$\pm$136 &  0.03$\pm$0.09 & -0.01$\pm$0.09\\
09:10:56.037 & 60:03:10.99 & 323.9 & 280 & 1227$\pm$163 &  76$\pm$129 &  0.01$\pm$0.09 & -0.01$\pm$0.10\\
\hline
&&&&Mask Z&&&\\
\hline
09:11:31.966 & 60:01:38.66 &  92.4 & 228 & 1305$\pm$ 11 & 140$\pm$ 15 &  0.15$\pm$0.05 &  0.02$\pm$0.13\\
09:11:28.467 & 60:02:51.46 &  98.9 & 298 & 1256$\pm$  9 & 157$\pm$ 12 &  0.08$\pm$0.05 & -0.01$\pm$0.07\\
09:11:35.293 & 60:02:03.77 &  30.6 & 234 & 1290$\pm$  5 & 181$\pm$  8 &  0.03$\pm$0.01 &  0.00$\pm$0.03\\
09:11:38.698 & 60:02:57.40 &  93.1 &  13 & 1357$\pm$ 14 & 138$\pm$ 26 & -0.02$\pm$0.07 & -0.06$\pm$0.07\\
09:11:25.602 & 60:01:31.53 & 135.6 & 244 & 1292$\pm$ 32 & 169$\pm$ 42 &  0.15$\pm$0.11 & -0.03$\pm$0.19\\
09:11:32.622 & 60:03:11.46 & 124.3 & 328 & 1301$\pm$ 20 & 148$\pm$ 26 &  0.00$\pm$0.06 &  0.02$\pm$0.06\\
09:11:39.464 & 60:01:47.13 &  61.5 & 151 & 1354$\pm$  5 & 143$\pm$  8 & -0.02$\pm$0.03 &  0.00$\pm$0.03\\
09:11:33.276 & 60:01:18.27 & 130.1 & 208 & 1328$\pm$ 20 & 140$\pm$ 26 & -0.02$\pm$0.06 & -0.07$\pm$0.07\\
09:11:37.061 & 60:01:52.67 &  49.2 & 186 & 1336$\pm$  5 & 154$\pm$  8 & -0.02$\pm$0.02 &  0.05$\pm$0.08\\
09:11:40.642 & 60:02:09.32 &  26.5 & 104 & 1426$\pm$  5 & 179$\pm$  8 & -0.02$\pm$0.01 & -0.02$\pm$0.01\\
09:11:35.163 & 60:03:04.31 & 106.5 & 341 & 1315$\pm$ 13 & 124$\pm$ 16 & -0.03$\pm$0.06 & -0.08$\pm$0.10\\
09:11:30.339 & 60:01:11.08 & 153.4 & 220 & 1310$\pm$ 25 & 140$\pm$ 33 &  0.06$\pm$0.08 & -0.09$\pm$0.17\\
09:11:35.188 & 60:02:16.80 &  16.7 & 276 & 1271$\pm$  5 & 196$\pm$  8 &  0.02$\pm$0.01 & -0.01$\pm$0.05\\
09:11:36.263 & 60:02:33.27 &  39.4 & 335 & 1311$\pm$  5 & 187$\pm$  8 &  0.02$\pm$0.01 &  0.02$\pm$0.02\\
09:11:36.492 & 60:02:45.93 &  66.5 & 348 & 1318$\pm$  5 & 153$\pm$ 11 &  0.02$\pm$0.03 &  0.07$\pm$0.06\\
09:11:28.833 & 60:01:58.16 &  77.2 & 255 & 1241$\pm$  8 & 148$\pm$  8 &  0.04$\pm$0.05 & -0.12$\pm$0.22\\
09:11:23.563 & 60:02:23.41 & 104.1 & 275 & 1153$\pm$ 12 & 106$\pm$ 23 &  0.11$\pm$0.12 & -0.02$\pm$0.25\\
09:11:51.197 & 60:01:24.50 & 144.5 & 116 & 1459$\pm$ 47 & 186$\pm$ 81 &  0.01$\pm$0.11 &  0.01$\pm$0.08\\
09:11:50.773 & 60:02:39.11 & 117.4 &  77 & 1466$\pm$ 36 & 158$\pm$ 62 & -0.05$\pm$0.12 &  0.02$\pm$0.12\\
09:11:48.366 & 60:03:22.15 & 173.5 &  51 & 1370$\pm$131 & 197$\pm$127 &  0.03$\pm$0.12 &  0.04$\pm$0.11\\
09:11:48.434 & 60:06:57.22 & 625.4 &  16 & 1464$\pm$189 & 461$\pm$161 & -0.12$\pm$0.11 &  0.01$\pm$0.14\\
\hline
\end{tabular}
\label{table:GC}
\end{center}
\end{table*}	

 \begin{table*}
\begin{center}
\contcaption{}
\begin{tabular}{cccccccc}
\hline
$\alpha$ & $\delta$ & $r$ & PA & $V_{\rm obs}$ & $\sigma$ & $h_3$ & $h_4$\\
(hh:mm:ss)&($^o$:':")&(arcsec)&(degree)&(km s$^{-1}$)&(km s$^{-1}$)&&\\
(1)&(2)&(3)&(4)&(5)&(6)&(7)&(8)\\
\hline
09:11:47.948 & 60:00:43.29 & 208.1 & 139 & 1409$\pm$165 & 147$\pm$125 & -0.09$\pm$0.14 &  0.02$\pm$0.13\\
09:11:46.978 & 60:03:32.05 & 187.7 &  43 & 1380$\pm$110 & 139$\pm$112 &  0.01$\pm$0.13 & -0.02$\pm$0.11\\
09:11:41.350 & 59:59:59.26 & 294.0 & 168 & 1351$\pm$173 & 163$\pm$112 & -0.02$\pm$0.06 &  0.01$\pm$0.07\\
09:11:38.908 & 59:59:13.61 & 393.5 & 176 & 1381$\pm$220 & 145$\pm$137 & -0.04$\pm$0.08 & -0.01$\pm$0.08\\
09:11:38.953 & 60:04:21.25 & 275.1 &   5 & 1297$\pm$193 & 166$\pm$126 & -0.05$\pm$0.08 & -0.01$\pm$0.07\\
09:11:34.378 & 60:04:34.25 & 300.8 & 351 & 1300$\pm$185 & 153$\pm$131 & -0.01$\pm$0.08 & -0.01$\pm$0.08\\
09:11:41.428 & 60:07:25.39 & 677.0 &   6 & 1221$\pm$211 & 190$\pm$147 & -0.05$\pm$0.10 & -0.06$\pm$0.10\\
09:11:26.564 & 60:12:28.29 & 326.6 & 352 & 1229$\pm$152 & 108$\pm$127 & -0.02$\pm$0.07 & -0.03$\pm$0.08\\
\hline
\end{tabular}
\end{center}
\end{table*}

\end{document}